\documentclass[%
floatfix,
 amsmath,amssymb,
 aps,longbibliography,
prb, twocolumn
]{revtex4-1}

\usepackage{graphicx,xcolor}% Include figure files
\usepackage{bm}% bold math
\usepackage{physics}
\usepackage{subcaption}
\usepackage{ragged2e}
\DeclareCaptionJustification{justified}{\justifying}
\usepackage{hyperref}
\usepackage{comment} % Comment several lines at once
\usepackage{orcidlink}
\usepackage[normalem]{ulem}
\begin{document}

\title{Current vortices in hexagonal graphene quantum dots}% Force line breaks with \\

\author{Eudes Gomes \orcidlink{0000-0002-3812-9331} }
\email{eudes.gomes@ufrpe.br}
\author{Fernando Moraes \orcidlink{0000-0001-7045-054X}}
 \email{fernando.jsmoraes@ufrpe.br}
\affiliation{ Departamento de F\'{\i}sica, Universidade Federal Rural de Pernambuco, 
 52171-900 Recife, PE, Brazil}

\date{\today}% It is always \today, today,

\begin{abstract}
Newly synthesized   nanostructures of graphene appear as a promising  breeding ground for new technology. Therefore, it is important to identify the role played by the boundary conditions  in their  electronic features. In this contribution we use the non-equilibrium Green's function  method coupled to tight-binding  theory to calculate and compare the  current patterns of  hexagonal graphene quantum dots,  with contacts placed at different edge locations. Our results reveal the formation of current vortices when the geometry of the contact position is not centrosymmetric. That is, when they are placed on non-frontal armchair-like vertices of the quantum dot. The introduction of defects, leading to disorder, significantly alters the current behavior, distorting bulk vortices and annihilating the edge currents. Nevertheless, periodically distributed defects on the edge allow for the appearance of current loops that dodge the defects. The presence of current vortices suggests the use of graphene quantum dots as nanomagnets or magnetic nanosensors.
\end{abstract}

\maketitle

%\section{\label{sec:level1}First-level heading}

\section{Introduction}
Graphene Quantum Dots (GQD) \cite{bacon2014graphene} or nanoflakes, have attracted a great deal of attention due to their possible electronic \cite{nishigaya2020graphene}, photonic \cite{choi2017unique} and medical \cite{chung2020photonic} applications. In this article, we investigate the possibility of generating current vortices in a hexagonal GQD. The shape choice is due to two reasons. First, the hexagonal shape preserves the local symmetry of the atomic network facilitating the calculations and second,   hexagonal nanographene has been recently successfully synthesized \cite{simpson2002synthesis,egberts2014frictional, lee2019synthesis}. The possibility of having current vortices in those nanoflakes suggests interesting applications, like their use as nanomagnets or magnetic nanosensors, for instance. This is the main motivation of this work. The electronic states of hexagonal and triangular GQDs have been thoroughly explored  by Zarenia and coworkers \cite{zarenia2011energy} and the dependence on size of the electronic structure of hexagonal GQDs was  studied by \cite{ghosh2017hexagonal} using density functional theory. 

Current vortices in graphene have been of great interest in recent years  \cite{levitov2016electron,danz2020vorticity}. They have been predicted to appear in the so-called hydrodynamic viscous regime, where electron-electron collisions are dominant (for a recent review see \cite{lucas2018hydrodynamics}) over electron-phonon or electron-impurity scattering. Since electron-phonon scattering is weak in graphene, as well as the degree of purity can be extremely high, the hydrodynamic behavior may be reached at reasonably low temperatures (say, above liquid nitrogen temperature). The backflow associated to the vortices in the hydrodynamic regime implies in a negative local resistance \cite{bandurin2016negative} giving indirect experimental evidence of their presence. This regime is not exclusive of graphene but may occur in ultra-clean 2D electron systems, as has been recently observed experimentally in a high purity GaAs quantum well \cite{levin2018vorticity}.  Although negative  resistances and current vortices seem to be  characteristics of the hydrodynamic regime, they may also appear in the ballistic regime, as recently verified by Gupta and coworkers \cite{gupta2021hydrodynamic}. 

\begin{figure}[htb!]
%\centering
\includegraphics[width=0.25\textwidth]{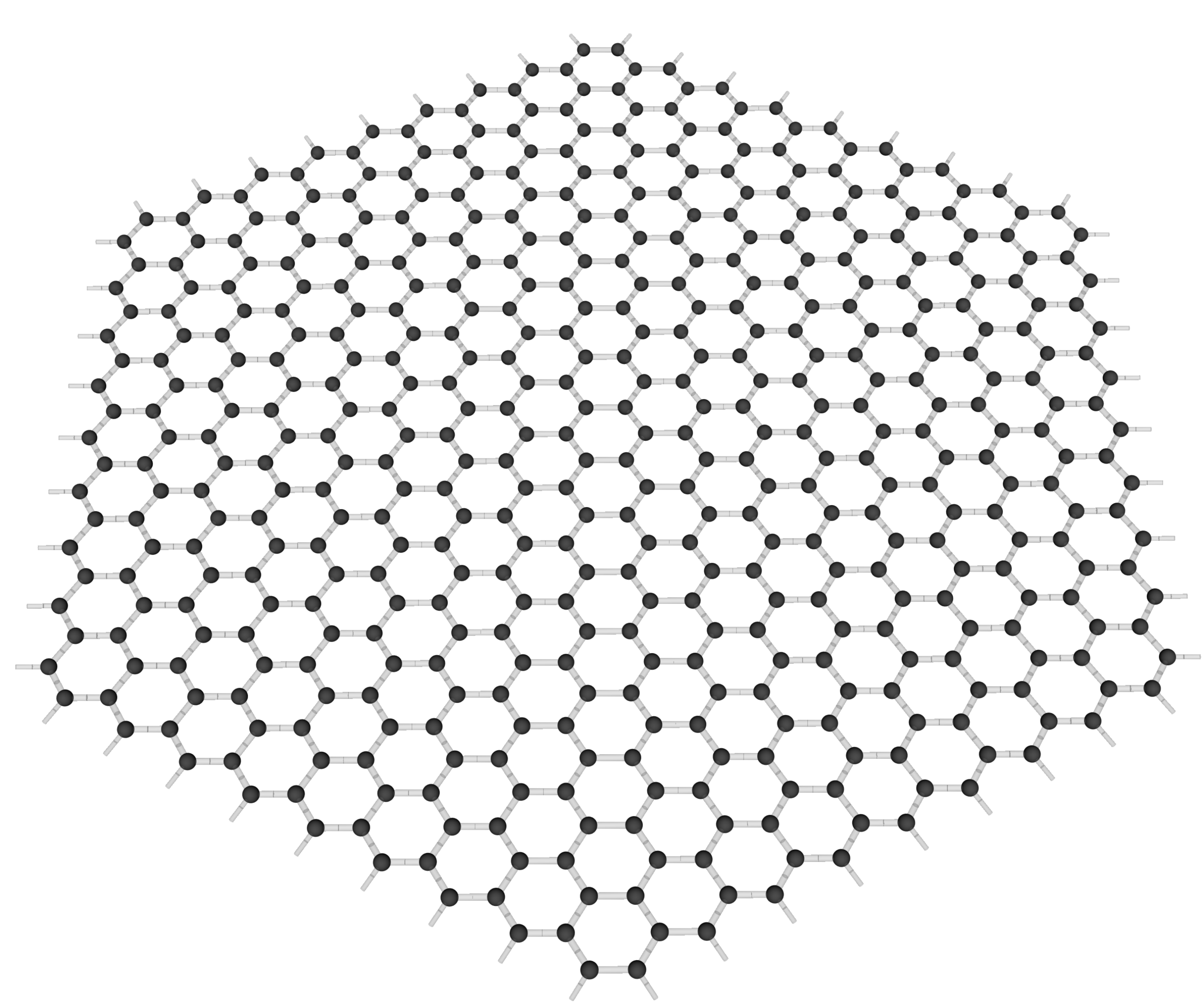}
\caption{Perspective view of a hexagonal graphene nanoflake $C_{486}H_{54}$. Hydrogen atoms (not shown) are attached to the atoms of the edge to avoid dangling bonds.}
\label{plane}
\end{figure}

Using a simple tight-binding model we show that, even without considering electron-electron interaction, current vortices may appear in a gated hexagonal graphene quantum dot with zigzag edge, depending on the contact locations and injection energy. We start by characterizing the elementary electronic properties, such as energy spectrum and density of states, of the  hexagonal structure $C_{486}H_{54}$ depicted in Fig. \ref{plane}. This is done by numerical diagonalization of the tight-binding Hamiltonian, and gives good agreement with the literature. The local density of states, transmission and local current are then obtained by coupling the tight-binding results to the Non-Equilibrium Green's Function (NEGF) method.  For selected contact configurations, we obtain  the transmission spectrum, which then allows us to choose the relevant values for the injection energy of each configuration. For some injection energies, vortices and their associated backflows appear clearly when the contacts are placed at two non-frontal armchair vertices.

Ring currents have  been found in an \textit{ab initio} study of a graphene nanoribbon  \cite{walz2014current}. Circular currents have also been found in hydrocarbon molecules, including coronene which is a smaller version of the system we study here, both by more robust techniques like Kohn-Sham Density Functional Theory coupled to NEGF \cite{stuyver2017exploring} and the simple tight-binding/NEGF \cite{ernzerhof2006electron} method we use here. This justifies \cite{stegmann2020current} the use of the latter approach in our studies, suggesting that the vortices we observe should also appear in more detailed treatments.

Although our model is oversimplified, it still seems to be accurate when describing effects due to defects on the edges \cite{aharon2021long}. The presence of these defects (defect here means removal of atoms) at the edges, and even in the bulk, were analyzed, and current vortices were found on some of our structures, depending on the degree of the disorder introduced by the defects. For a detailed discussion on the role played by defects on the edges of graphene nanoribbons see \cite{li2008quantum}. 

A note on the nomenclature used for closed current loops is due here. While in a superfluid one might have a free vortex, that needs not an energy supply to be maintained, in this work we use the term ``vortex'' to designate a forced vortex, in the sense that a potential difference is needed to keep it, due to ohmic losses. These are like the forced vortices that appear upon stirring water in a kitchen pot. On the other hand, in superconducting graphene, for instance, one might have free vortices, which will persist after turning off the potential difference that originated them. 
 
\section{Methods}

\subsection{Tight-binding calculations}
The tight-binding approach for molecules uses the most relevant atomic orbitals  $\lambda $ associated to each atom $i$ as a basis to construct the single particle state $\ket{\psi}=\sum\limits_{i\lambda }\ket{i \lambda }$. In our calculations we will use only  $\pi$ orbitals  since these are mostly responsible for the transport properties in graphene. The tight-binding Hamiltonian, written in the  local $\pi$ orbitals' basis, is then given by
\begin{equation}
    H= \sum_i \varepsilon_i \ket{i}\bra{i} + \frac{1}{2}\sum_{i\neq j}t_{ij}\ket{i}\bra{j}, \label{hamiltonian}
\end{equation}
where $\varepsilon_i$ is the on-site orbital energy and $t_{ij}$ is the hopping integral. Considering only first-neighbor interactions, we define the overlap matrix
 \[
    S_{ij}= 
\begin{cases}
    1,& \text{if } i = j\\
    s,              & \text{if i, j are neighbors} \\
    0, & \text{otherwise}
\end{cases}
\]
 and the hopping matrix
 \[
    H_{ij}= 
\begin{cases}
    \epsilon_{i},& \text{if } i = j\\
    t,              & \text{if i, j are neighbors} \\
    0, & \text{otherwise}
\end{cases}
\]
 and use $\epsilon_{i} = 0$, $t = -3.1\, eV$ and $s = 0.12$,  following Ref.\cite{dresselhaus1998physical}. Our task now reduces to finding the eigenvalues $E$ of the Hamiltonian \eqref{hamiltonian} by solving the characteristic equation
 \begin{equation}
     \text{det}(H-E S)=0 .
     \label{sec}
 \end{equation}
 The hopping and overlap matrices  $(H,S)$ are easily constructed when one knows the connections among the atoms. In our case, each carbon atom is connected to another three except at the edge where the connectivity is two. 

\subsection{The Non-Equilibrium Green's Function Method}
\begin{figure}[ht]
    %\centering
    \includegraphics[width=0.34\textwidth]{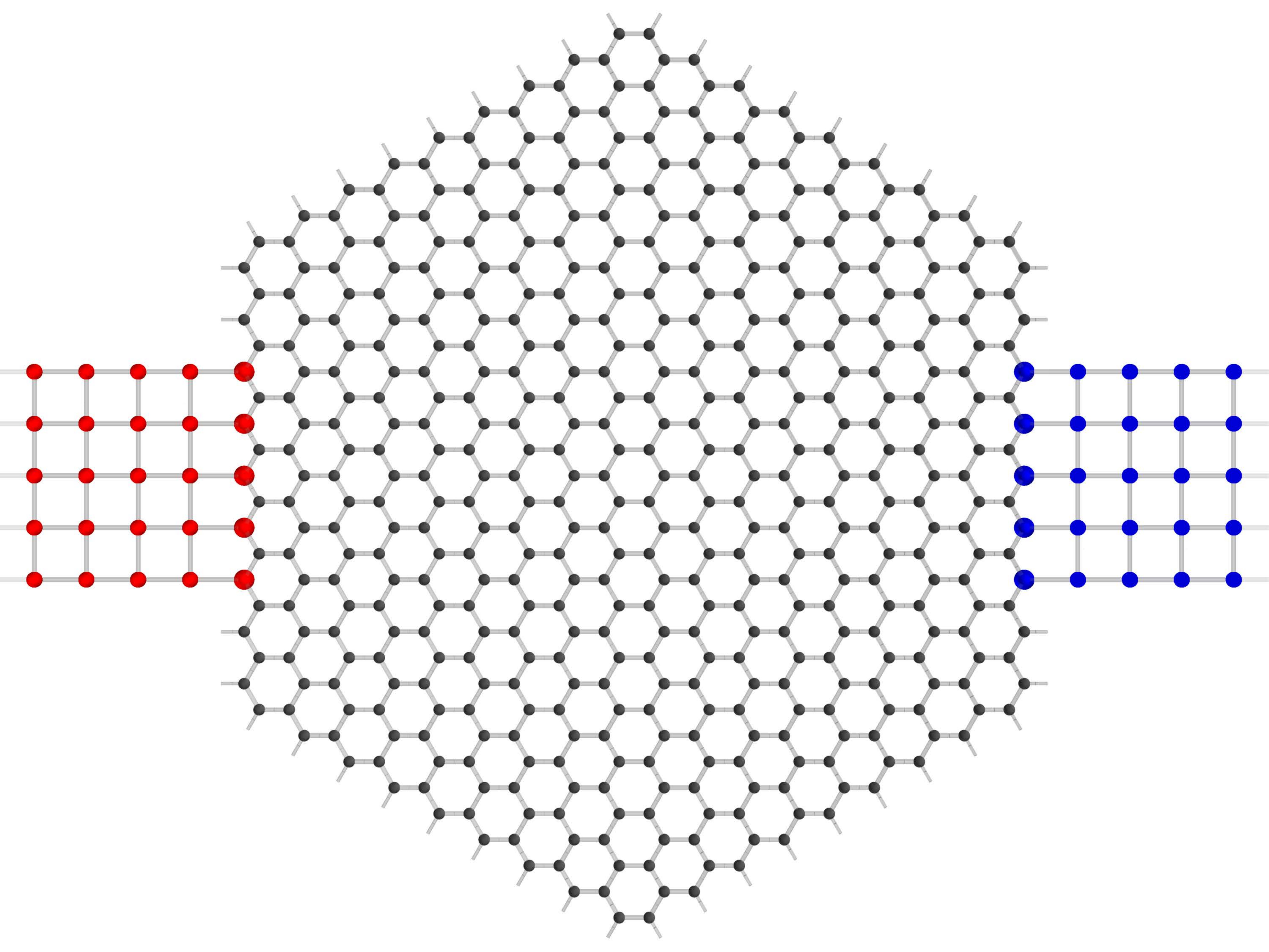}
    \caption{Source (red) and drain (blue)   contacts at the hexagonal GQD edge.}
    \label{dotsleads}
\end{figure}

In order to obtain the transport properties of the nanoflakes, we resort to the  Non-Equilibrium Green's Function (NEGF) method which, coupled to the tight-binding formalism, is well known as an excellent tool to study current flow and transport phenomena in nanostructures\cite{stegmann2016current,stegmann2018current,stegmann2020current,datta2000nanoscale,schomerus2007effective,dubois2009electronic}.
To use the NEGF method we need to attach to the nanostructure two or more leads, which are metallic contacts acting as reservoirs. These contacts are modeled using a square lattice (see Fig. \ref{dotsleads}) and the reservoirs are kept at different chemical potentials, keeping the whole system out of equilibrium. Each reservoir is characterized by its own Fermi distribution. We can use  NEGF to calculate  electronic  properties such as, the local current $I_{ij}$ between neighboring lattice sites $i$ and $j$, the transmission,  and the local density of states (LDOS). 
Since the NEGF method is widely discussed in\cite{datta1997electronic,datta2005quantum}, we summarize here only the essential concepts. 
We start by writing down the Green's function of the system, given by
\begin{equation}
    G(E) = (ES - H - \Sigma_{i})^{-1},
    \label{green}
\end{equation}
where E is the single-particle energy of the injected electrons and S, H and $\Sigma_{i}$ are the overlap, hopping and the so called self-energies, respectively. The self-energies account  for the  contacts attached to the nanoflakes, which are modeled by a semi-infinite square lattice as described in \cite{lewenkopf2013recursive,stegmann2014quantum}. Besides the real contacts associated to the reservoirs, it is also possible to attach virtual contacts to the remaining edge atoms. These contacts can be placed in order to suppress boundary effects by means of an imaginary self-energy. Here, these virtual contacts are modeled using  the wide-band approximation, which can be expressed as $\Sigma^{edge} = -i\nu\sum_{n} \ket{n}\bra{n}$, where $\nu>0$ is a parameter \cite{stegmann2016current,stegmann2013magnetotransport}. Later on, in Subsection \ref{edge effects}, we shall show the role played by these virtual contacts on the transmission. In Fig. \ref{dotsleads} we show the hexagonal GQD of interest with source (red) and  drain (blue) contacts,  with a few sites of the respective square lattices.  The injection is modeled by an inscattering function of the form 
\begin{equation}
    \Sigma^{in} = \Gamma_{s} f_{s} + \Gamma_{d} f_{d},
\end{equation}
where subscripts $s (d)$ stand for source (drain) and the function $\Gamma_{k}=i(\Sigma_{k} - \Sigma_{k}^{\dagger})$, $k = s,d$, is called broadening matrix. The functions $f_{s,d}$ are Fermi distributions for the reservoirs.
With this formalism, we obtain the transport properties by means of the transmission function\cite{stegmann2020current}
\begin{equation}
    T(E) = \Tr[\Im(\Sigma_{s})G \Im(\Sigma_{d})G^{\dagger}], \label{transm}
\end{equation}
as well as the local current between atoms $i$ and $j$, 
\begin{equation}
    I(E)_{ij} = \frac{2e}{h}\Im(H_{ij}^{*}G_{ij}^{n}),
    \label{localtrans}
\end{equation}
where $G^{n}=G\Sigma^{in}G^{\dagger}$ is the correlation function\cite{datta2005quantum,stegmann2016current}. Another important information that we can obtain from the Green's function method is the local density of states, which is given by\cite{datta1997electronic}
\begin{equation}
    \rho(E,i) = -\frac{1}{\pi}\Im(G_{ii}(E)). \label{ldos}
\end{equation}
When studying  the nanoflake without  contacts, we  write the Green's function as
\begin{equation}
    G(E) = \left[(E+ i \mu) S-H \right]^{-1}, \label{Gnocontact}
\end{equation}
where $\mu > 0$ is a parameter and $i = \sqrt{-1}$.
All relevant results in this work were obtained by means of the above equations coupled to the tight-binding model described in the previous subsection. We used the value $\mu = 0.0025\, eV$. For more information on the role of the parameter, $\mu$ see \cite{schmitteckert2010calculating}. Equation \eqref{Gnocontact} will be used to evaluate the LDOS without any influence of the contacts. To evaluate the LDOS taking into account  the presence of the contacts, we can use the complete Green's equation given by \eqref{green}. The code to perform the calculations  was written in Python, and for the visualization of  the LDOS and current patterns we used Mayavi \cite{mayavi}.
\section{Results}
\subsection{Energy spectrum and density of states}
As a verification of our method, we performed the electronic characterization of the  hexagonal GQD and compared it to known results for similar structures. 
\begin{figure}[htp!]
	\centering
	 \includegraphics[width=0.9\columnwidth]{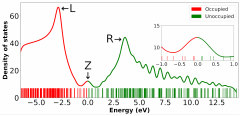}
	\caption{Density of States and energy spectrum for the hexagonal graphene quantum dot. Red (green)  indicates  occupied (unoccupied) states. The letters L,R,Z indicate the position of the two van Hove  and zero states, respectively. The corresponding energy values are   $L = -2.8685\, eV$, $R = 3.6014\, eV$ and $Z = 0.0\, eV$. The inset shows the states around Z.}
	\label{dispersio+DOS}
\end{figure}
In Fig. \ref{dispersio+DOS} we show the energy spectrum, as obtained from the diagonalization of Eq. \eqref{sec},  and the corresponding Gaussian broadened (full width at half maximum $=\,0.3\, eV$) Density of States (DOS). The overall shape and major peaks of the DOS, correspond to the ones obtained previously in the literature\cite{stein1987pi,ulloa2013cone} for comparable structures. Note  the presence of the peak at zero energy  in the DOS (indicated by  Z in Fig. \ref{dispersio+DOS}). 
\begin{figure}[htp!]
	%\centering
	 \includegraphics[width=0.9\columnwidth]{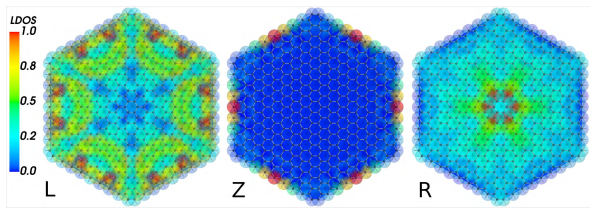}
	\caption{Local Density of States corresponding to the three selected DOS peaks in Fig. \ref{dispersio+DOS}.  Note the edge states in the central illustration (Z) corresponding to the zero modes. }
	\label{three_ldos}
\end{figure}
As it is well known\cite{wimmer2010robustness},  zero modes in 2D carbon structures correspond to edge states, as confirmed by the map of the Local Density of States (LDOS) at zero energy (Fig. \ref{three_ldos} Z) on the structure. This is in close agreement with the results of Ref.\cite{stein1987pi} for the Highest Occupied Molecular Orbital (HOMO) and of Ref.\cite{ghosh2017hexagonal} for the Lowest Unoccupied Molecular orbital (LUMO) of zigzag hexagonal carbon nanoflakes. To be precise, there is no zero energy state for this structure but the HOMO and LUMO states, which are symmetrically placed around zero, with a gap of nearly $0.2\, eV$. As verified in Ref.\cite{ghosh2017hexagonal},  increasing   the size of the structure  reduces the gap, eventually leading to its closure in   the infinite graphene limit. 
The LDOS at the van Hove peaks reflect  the common symmetries of the hexagonal lattice and the edge shape, as can be seen in Fig. \ref{three_ldos}, for the first (left) and for the second (right) peak. 
A detailed study of the effects on the van Hove peaks  due to defects or edge shape in flat graphene nanoflakes can be found in Ref.\cite{zhou2012van} and the  consequences for optical properties in\cite{pohle2018symmetry}.

\subsection{Transmission and current}
The total transmission as a function of the energy can be obtained from equation \eqref{transm}, but note that it also depends on the leads configuration. 
%\begin{figure}[ht]
%    \centering
%    \includegraphics[width=0.45\textwidth]{current_0p6769n_0p0253.pdf}
%    \caption{Top: transmission as a function of injection energy for the contact configuration shown in Fig. \ref{dotsleads}. Bottom: current patterns corresponding to the selected transmission peaks L (left) and R (right). The color bar presents a scale for the current intensity in units of $\frac{2e}{h}$. Contacts are represented by colored dots: source (red) and drain (blue). }    \label{4contact}
%\end{figure}
%\begin{figure}[htbp]
%    \centering
%    \includegraphics[width=0.45\textwidth]{current_0p0163n_0p3458.pdf}
%    \caption{Top: transmission as a function of injection energy for contacts placed on two non-adjacent edges of the hexagonal GQD. Bottom: current patterns corresponding to the selected transmission peaks L (left) and R (right). The color bar presents a scale for the current intensity in units of $\frac{2e}{h}$. Contacts are represented by colored dots: source (red) and drain (blue). } \label{5contact}
%\end{figure}

%%%%%%%%%%%%%%%% O Código comentado acima retorna ao original

\begin{figure}[htp!]
\begin{subfigure}[b]{0.43\textwidth}

 \includegraphics[width=1\textwidth]{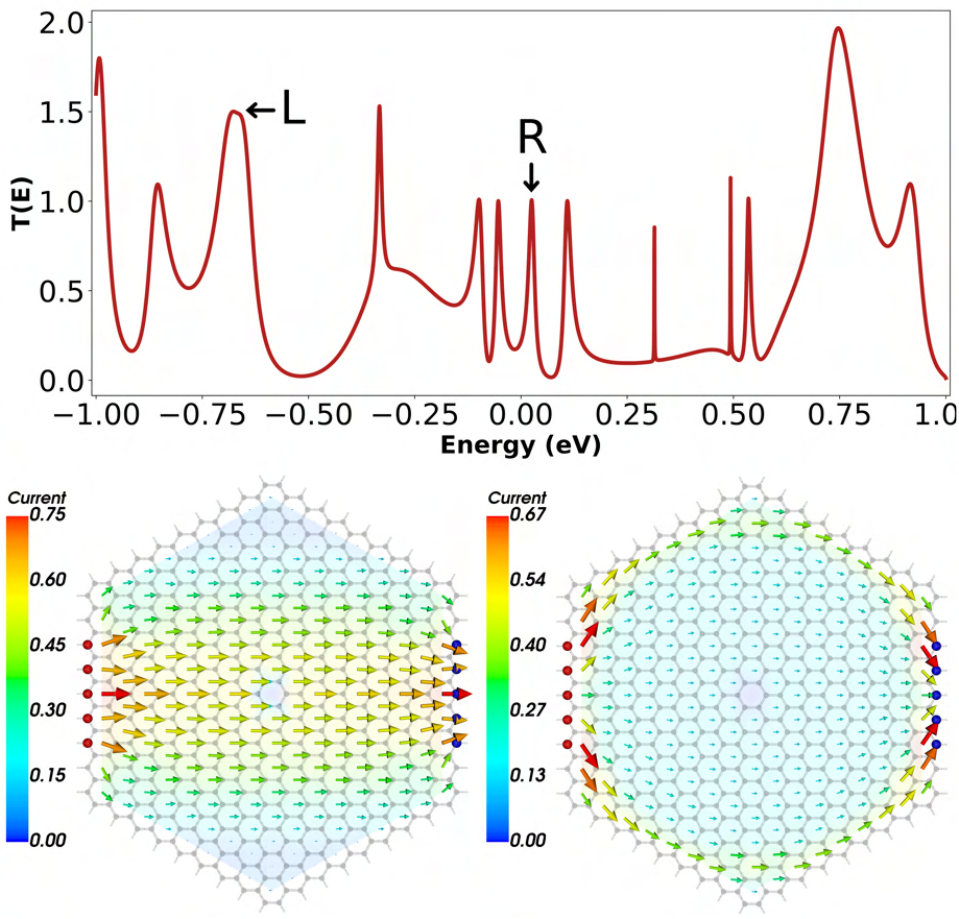}
  \caption{Top: transmission as a function of injection energy for the contact configuration shown in Fig. \ref{dotsleads}. }
  \label{4contact}
\end{subfigure}
\begin{subfigure}[b]{0.43\textwidth}
  \includegraphics[width=1\textwidth]{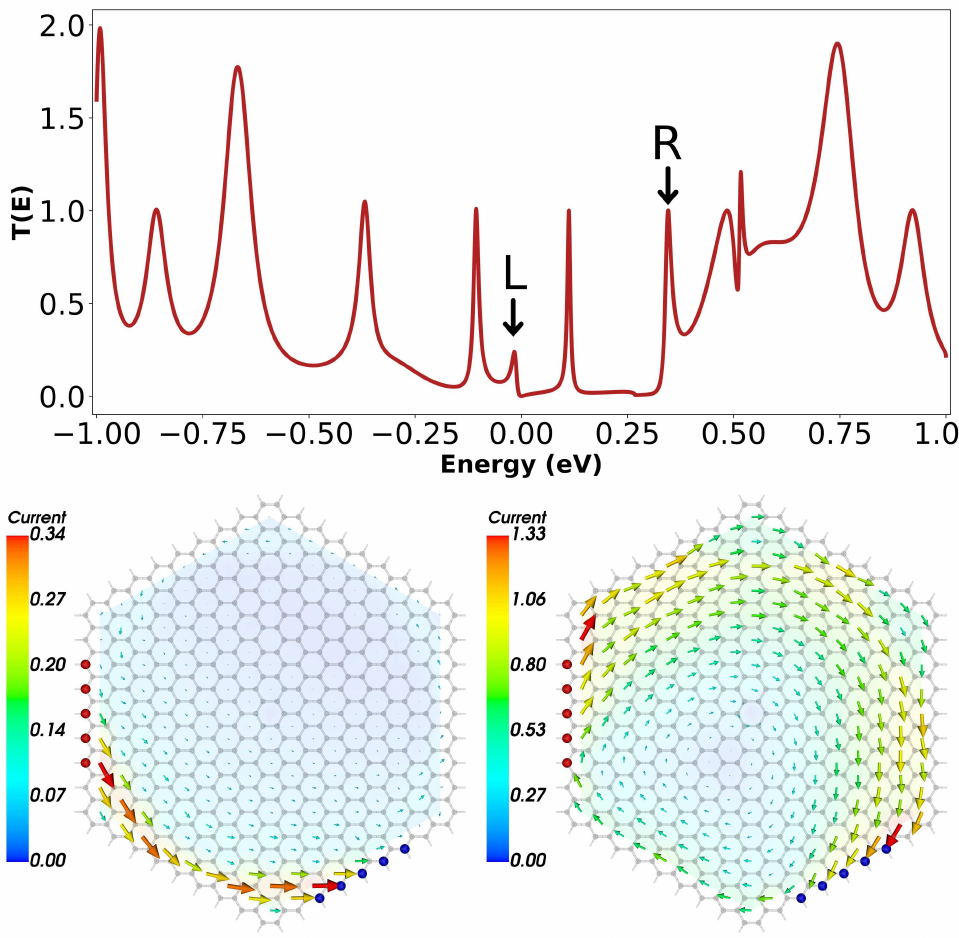}
  \caption{Top: transmission as a function of injection energy for contacts placed on two non-adjacent edges of the hexagonal GQD.}
  \label{5contact}
\end{subfigure}
\caption{Bottom of each figure: current patterns corresponding to the selected transmission peaks, L (left) and R (right). The color bar presents a scale for the current intensity in units of $\frac{2e}{h}$. Contacts are represented by colored dots: source (red) and drain (blue). }
\end{figure}

We start by analyzing the  configuration  presented in Fig. \ref{dotsleads}, with contacts placed on opposing edges. In the upper part of Fig. \ref{4contact} we show the transmission as function of injection energy for this configuration. In the lower part, we show the current patterns for two selected values of injection energy. A note on the current diagrams: each arrow, except those in the central polygon when they appear, is the resulting sum over the currents at the edges of each polygon in the structure. The current patterns shown in Fig. \ref{4contact} are representative of what is obtained at the other transmission peaks: the current leaves the source and goes to the drain with some lateral spreading whose degree varies with the injection energy. The closer the injection energy is to zero, the greater the spreading until the current goes mostly through the edge states (compare with Fig. \ref{three_ldos} Z). Obviously, the symmetric arrangement of the  contacts across the sample does not favor a circular current pattern  formation. As it will be seen next, this  happens when the  contacts are placed asymmetrically across the flake.
\begin{figure}[h!]
 \centering
 \includegraphics[width=0.45\textwidth]{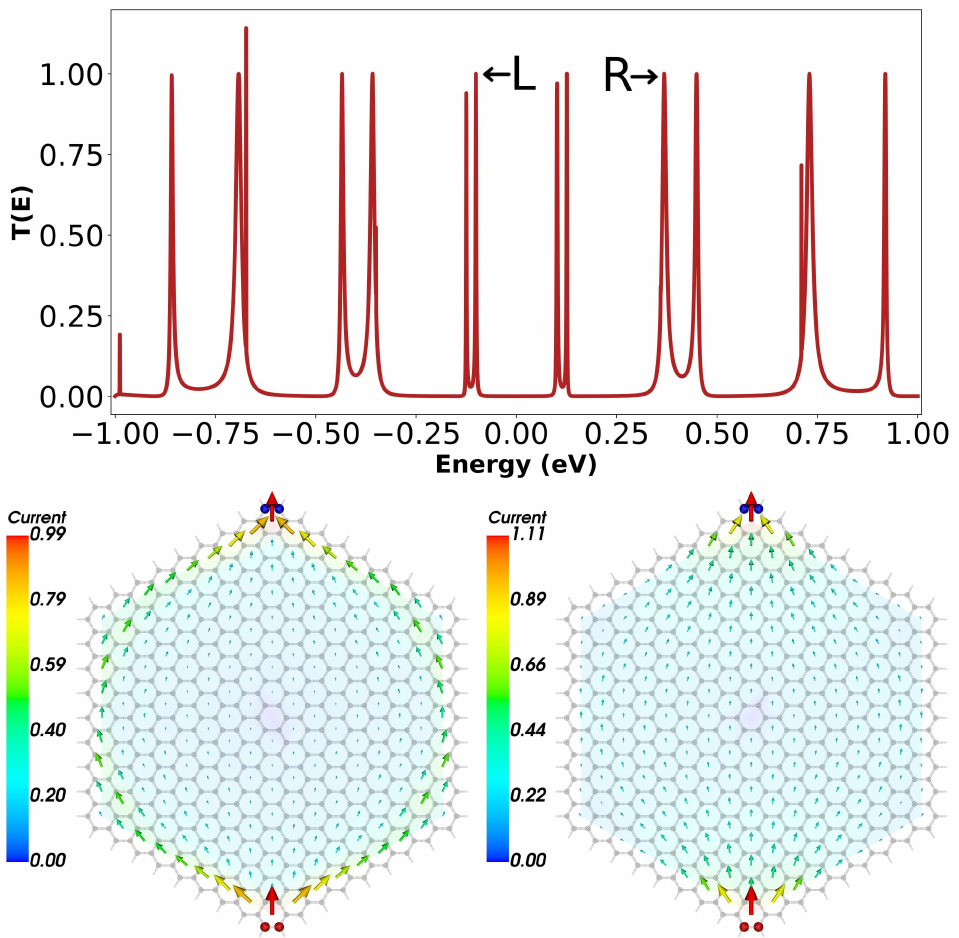}
\caption{Top: transmission as a function of injection energy for contacts placed at opposing vertices.  Bottom: current patterns corresponding to the selected transmission peaks L (left) and R (right). The color bar presents a scale for the current intensity in units of $\frac{2e}{h}$.  \label{plane_current_front} Contacts are represented by colored dots: source (red) and drain (blue).}
\label{2atomfront}
\end{figure}

As another example, we show the results for contacts placed on non-frontal edges of the hexagonal GQD. In Fig. \ref{5contact} we show the transmission as a function of injection energy and two selected current patterns, corresponding to the transmission peaks indicated. Again, we see that the current follows the path of the edge states for a choice of injection energy near zero (figure on the left). An  unevenly distributed vortex-like current pattern appears in the figure on the right.

Now, we investigate the case of only two contacts placed at the carbon atoms in opposing vertices, where the edge configuration is armchair-like. Later on, in Subsection \ref{subB}, we will analyze the zigzag case. As in the previous case, in the lower part of Fig. \ref{plane_current_front} we show two representative current patterns corresponding to the selected transmission peaks shown in the upper part of the figure. As before, we see that the current, for energies near zero, follows the path of the edge states. Otherwise, it tends to spread out all over the structure.  Note the mirror symmetry of both the L and R current patterns upon reflection by the axis joining the opposing contacts. Again, the symmetry of the contacts across the flake does not favor vortex formation.
\begin{figure}[htbp]
 \centering
  \includegraphics[width=0.45\textwidth]{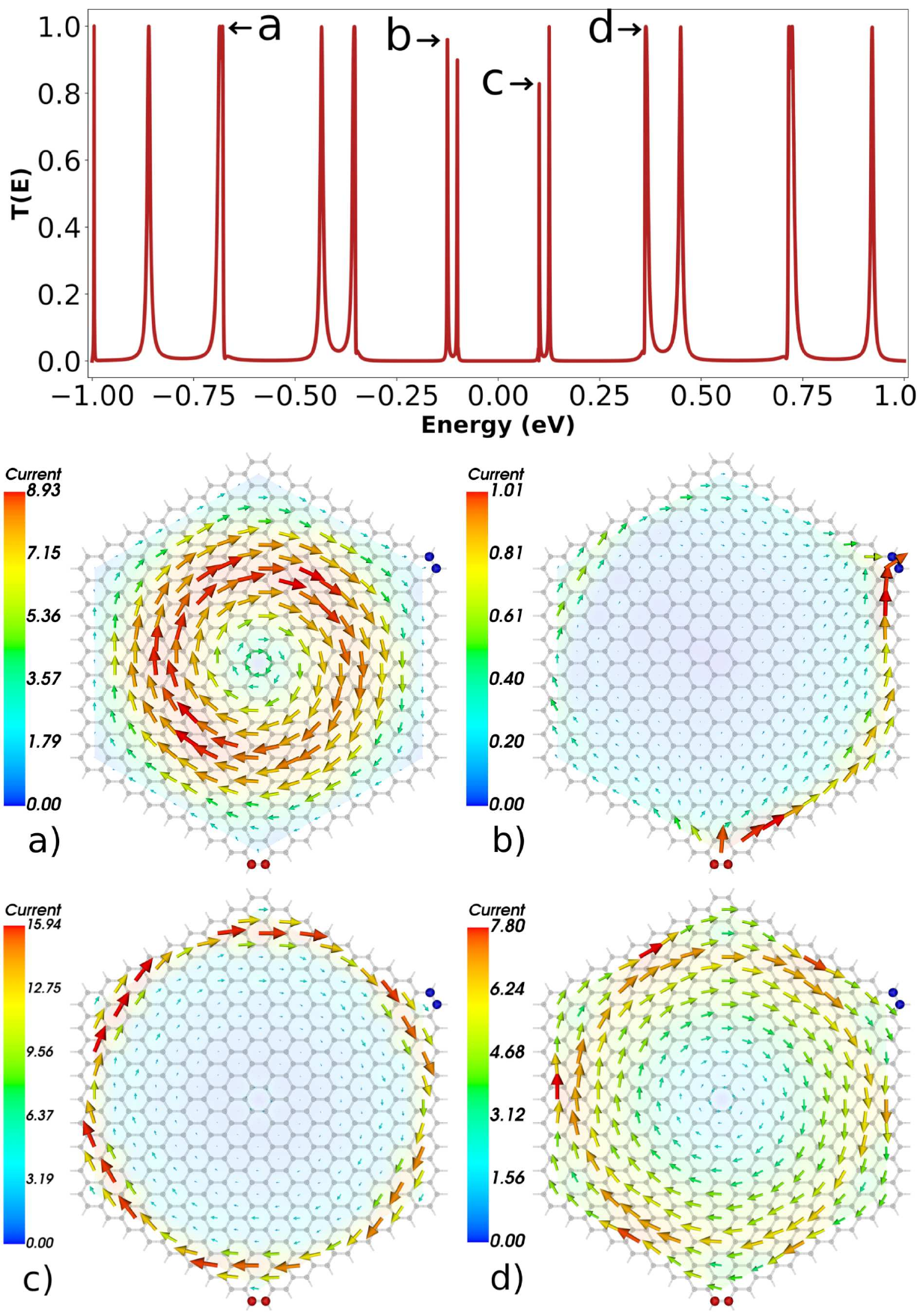}
 \caption{Top: transmission as a function of injection energy for contacts placed at two vertices.  Bottom: current patterns corresponding to the selected transmission peaks in the sequence a, b, c, d, from left to right, from top to bottom. The color bar presents a scale for the current intensity in units of $\frac{2e}{h}$. Contacts are represented by colored dots: source (red) and drain (blue).} \label{vortfig}
\end{figure} 

\begin{figure}[htbp]
 \centering
  \includegraphics[width=0.45\textwidth]{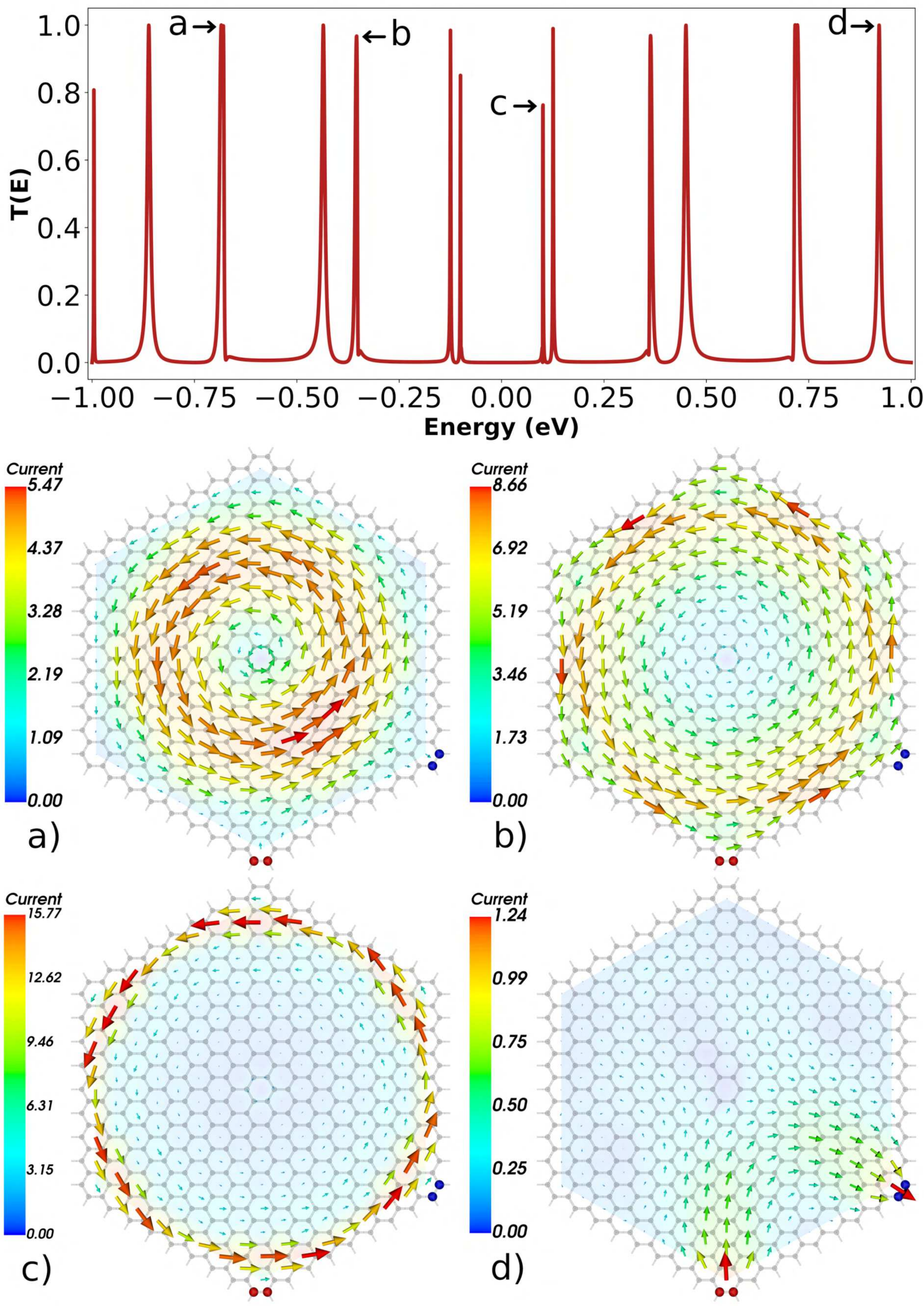}
 \caption{ Top: transmission as a function of injection energy for contacts placed at two adjacent armchair-like vertices.  Bottom: current patterns corresponding to the selected transmission peaks in the sequence a, b, c, d, from left to right, from top to bottom. The color bar presents a scale for the current intensity in units of $\frac{2e}{h}$. Contacts are represented by colored dots: source (red) and drain (blue). } \label{vortfig2}
\end{figure}

Next, we investigate the case of non-frontal vertex contacts placed across the hexagonal GQD. In the upper part of Fig. \ref{vortfig} it is shown the transmission for this configuration as a function of injection energy. There,  the peaks chosen for the corresponding current patterns depicted in the lower part of the figure, are indicated. Now, induced by both the non-frontal position of the contacts and the chosen energy, vortices, as well as circular edge current, appear. Corresponding to the near zero energy peaks b and c we see, once again, the prevalence of the edge states guiding the current flow. Note, however, that only in Fig. \ref{vortfig}c there exists a circular edge current. Comparing Figs. \ref{2atomfront}a and \ref{vortfig}b we see that, while in the former the current splits equally over the two equivalent paths along the edge, it chooses the shorter path in the latter.

Another case, now with contacts placed at two vertices sharing the same edge, is shown in Fig. \ref{vortfig2}. Again,  both vortices and circular edge current, appear. As before, the closer the injection energy is to zero, the closer the current path to the edge. It is worth commenting on the Fig. \ref{vortfig2}d: note that its injection energy is quite high, away from the small range of energy around zero that sustains  edge states. Therefore, the current does not flow directly through the edge, it starts on a radial path instead, until it is attracted by the chemical potential of the second reservoir. We note that in all cases studied, the transmission peaks are of the same order of magnitude. Nevertheless, the corresponding maximum currents may change by an order of magnitude for different injection energies. In particular, the maximum current is higher whenever a vortex appears. This apparent disagreement between transmission and maximum current is due to the fact that the transmission is related to the actual current that reaches the drain, while the maximum current is local. In other words, when a vortex is formed, a great part of the current is captured and remains circulating while only a fraction contributes to the transmission. This behavior has also been observed in current vortices in graphene nanoribbons \cite{walz2014current}. Another way of looking at this, is to observe that the vortex is a composition of an incoming and an outgoing current, which therefore reduce the net current reaching the electrodes. According to Ref. \cite{stegmann2020current}, the presence of the backflow is due to quantum interference between nearly degenerate states mediated by the off-diagonal terms of the complete Hamiltonian. Also, two-atom wide contacts can only sustain a few propagating modes, which in turn limit the maximum reached by the transmission. In fact, for a perfect system, the transmission is quantized, and it grows with the width of the contacts. More on this in Sect. \ref{subB}.

\section{Discussion \label{discuss} }

\subsection{Edge effects \label{edge effects} }

It is well known that the edges play an important role in the electronic properties of graphene. In graphene nanoribbons there are two possible edge types: armchair or zigzag. These conformations can also be found in other carbon-based materials such as cis-polyacetylene (armchair), trans-polyacetylene (zigzag) or nanotubes \cite{dresselhaus1998physical}. Although the bulk configuration of a graphene nanoribbon may be similar to the graphene sheet, irrespective of its edge, it is on the edge that we can find  interesting electronic properties \cite{malysheva2008spectrum}. On one hand, the zigzag edge type is known to be metallic while, on the other hand, the armchair edge type is known to be either metallic or semiconducting, depending on the width of the ribbon \cite{wakabayashi2010electronic}. Notice that in the cases where the current is circulating strictly on the edges, as in Figs. \ref{vortfig}c and \ref{vortfig2}c, the maximum of the current is considerably higher than those corresponding to other contact configurations and energy peaks. This is so, due to the metallic characteristic of the zigzag edges. In fact, it was shown by Ref. \cite{mabillard2014spatial} that at low energy the current flow is localized on the zigzag edges states, in the same sense as those depicted in Figs. \ref{vortfig}c and \ref{vortfig2}c, which may explain the increase of the maximum of the current.  Furthermore, the role of the edges states  has been widely studied in several contexts \cite{uri2020nanoscale,liu2017helical,ji2013influence,zhu2017edge,christensen2014classical,pramanik2014effect,fujii2014role,stegmann2013magnetotransport}. Nonetheless, both armchair and zigzag edges have their importance in magnetic effects and  can be used in spintronic devices as well. See, for instance, \cite{farghadan2013magnetic,farghadan2012spin,saffarzadeh2011spin,farghadan2016magnetism} for a detailed description. 
\begin{figure}[htp!]
	\centering
	 \includegraphics[width=1\columnwidth]{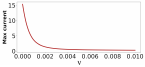}
	\caption{Maximum current as a function of the parameter $\nu$ for the case depicted in Fig. \ref{vortfig}c. The current is given in units of $2e/h$ and the parameter $\nu$  in eV.    }
	\label{maxvsnu}
\end{figure}

It is also possible  to suppress the edge effects by including virtual reservoirs\cite{stegmann2016current} in the calculations. These  are modeled using the wide-band approximation. Actually, this model is quite accurate when describing metallic contacts such as gold (for more information about the wide-band approximation, see \cite{verzijl2013applicability}). Figure \ref{maxvsnu} shows the maximum current of Fig. \ref{vortfig}c as a function of the parameter $\nu$, and we can clearly see the effect of the virtual contacts, which act like absorbing  boundaries, mimicking inﬁnite dimensions of the graphene sheet \cite{stegmann2016current}, reducing drastically the intensity of the current and nearly extinguishing the circular flow. Naturally, the same happens in the case of Fig. \ref{vortfig2}c.

%%%%%%%%%%%

\subsection{Effects of the contacts \label{subB}}

In the previous section, we presented a few different cases of contact geometry on the hexagonal  GQD. In each case, we used  the transmission as a function of injection energy as a guide to choose the energies where there is current flowing between the contacts. Quite naturally, the complexity of the transmission profile increases with the number of contacts {(see Figs. \ref{4contact} and \ref{5contact})}  while fewer contacts lead to sharper transmission peaks (see Figs. \ref{plane_current_front}, \ref{vortfig} and \ref{vortfig2}). This is  due to the width of the reservoirs. In the case where the contact  is wider, say, Figs. \ref{4contact} and \ref{5contact}, the transmission becomes smoother. At the contacts, the current would, in principle, be carried   by an infinite number of transverse modes  but, due to the limited size of the  interface reservoir-GQD, only a few of these enter into the device. This is so, due to the interface resistance (see chapter 2 of Ref.\cite{datta1997electronic}). 
%Trocar 10 por 11
%\begin{figure}[htp!]
%	\centering
%	 \includegraphics[width=1\columnwidth]{current_h_1p3063n_1p6684n.pdf}
%	\caption{Multiple-vortex patterns for two-atom-wide-contact configurations. The energies of both left and right figures are $-1.3063eV$ and $-1.6680eV$, respectively. These energies are outside the energy range showed in the inset of Fig. \ref{dispersio+DOS}, and still  form interesting current patterns. The color bar presents a scale for the current intensity in units of $\frac{2e}{h}$. Contacts are represented by colored dots: source (red) and drain (blue).}
%	\label{maxvsnu1}
%\end{figure}
% essa é a 11 original
\begin{figure}[htbp]
    \centering
    \includegraphics[width=0.45\textwidth]{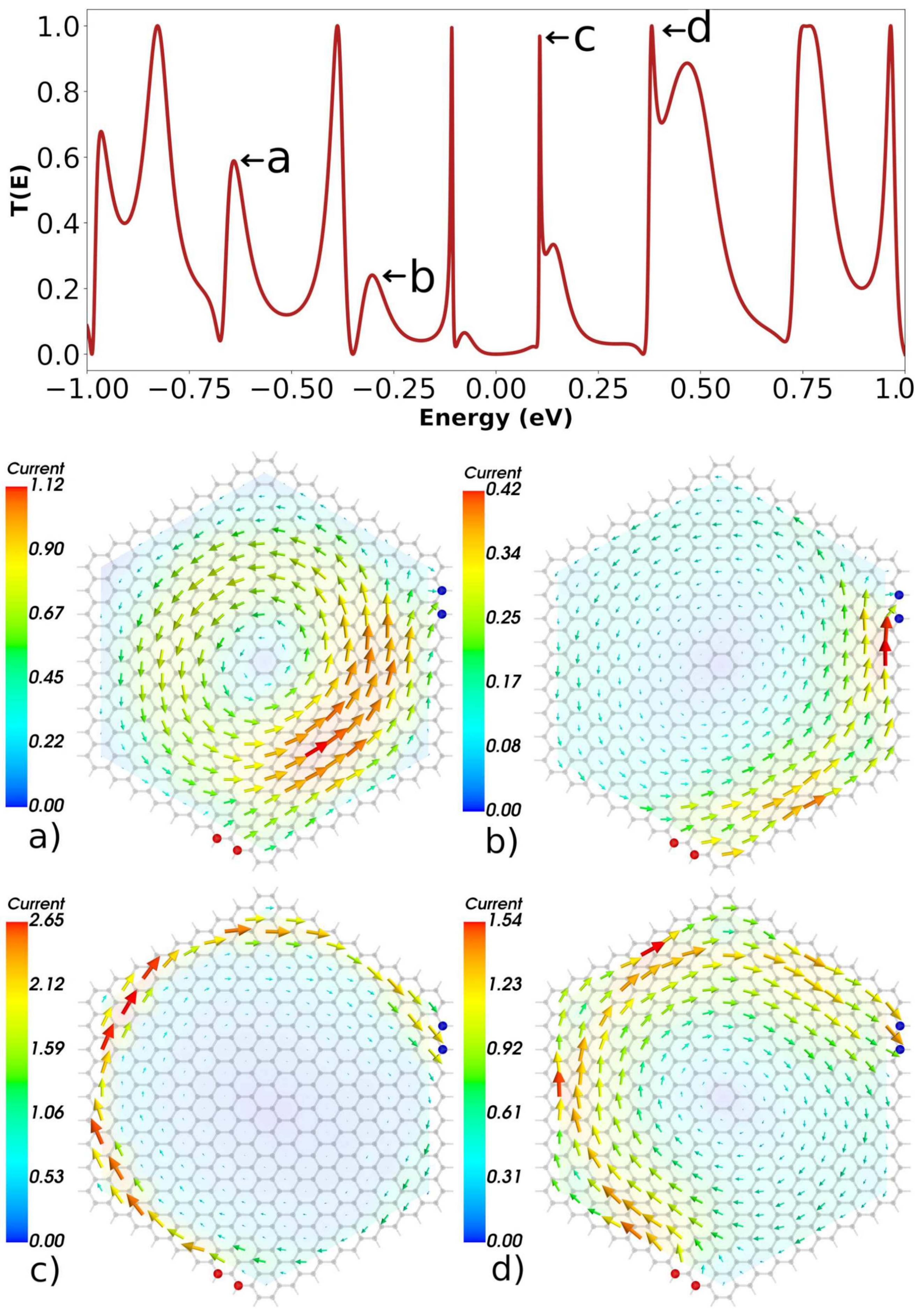}
    \caption{Top: transmission as a function of injection energy for contacts asymmetrically placed on the zigzag edge of the hexagonal GQD. Bottom: current patterns corresponding to the selected transmission peaks in the sequence a, b, c, d, from left to right, from top to bottom. The color bar presents a scale for the current intensity in units of $\frac{2e}{h}$. Contacts are represented by colored dots: source (red) and drain (blue). } \label{2zigzag}
\end{figure}

Also, it may be noticed that the maximum of the transmission in the energy range used in this work is greater for those structures with wider contacts. Again, this is due to the fact that wider contacts can carry more propagating modes. Actually, in an ideal system, the maximum number of propagating modes is proportional to the width of the contacts which leads to a quantized conductance \cite{datta1997electronic,stegmann2014quantum}. When dealing with the 2-atom-wide contacts, which are the cases of Figs. \ref{2atomfront}, \ref{vortfig}, \ref{2zigzag} and \ref{vortfig2}, the transmission shows sharper peaks, which are a result of the short energy range where resonant states can appear. It is worth  mentioning that our square reservoirs can be seen as being composed of individual 1D chains and the transmission of each chain is only perfected in its conduction band, vanishing outside it.   In addition, the geometry and the edge type have also a great influence on the conductance as shown by \cite{torres2018tuning}, when studying three-terminal systems. The break in the stepped-like conductance was also shown to appear in graphene nanoribbons with defects, both for zigzag and armchair edges \cite{gorjizadeh2008effects} and in disordered graphene nanoribbons\ \cite{mucciolo2009conductance}. Notice that, for the cases of 2-atom-wide contacts, the transmission peaks are quasi-symmetrical with respect to  zero energy and some of these symmetrical spikes give rise to similar current patterns.
% essa é a 10 original
\begin{figure}[htp!]
	\centering
	 \includegraphics[width=1\columnwidth]{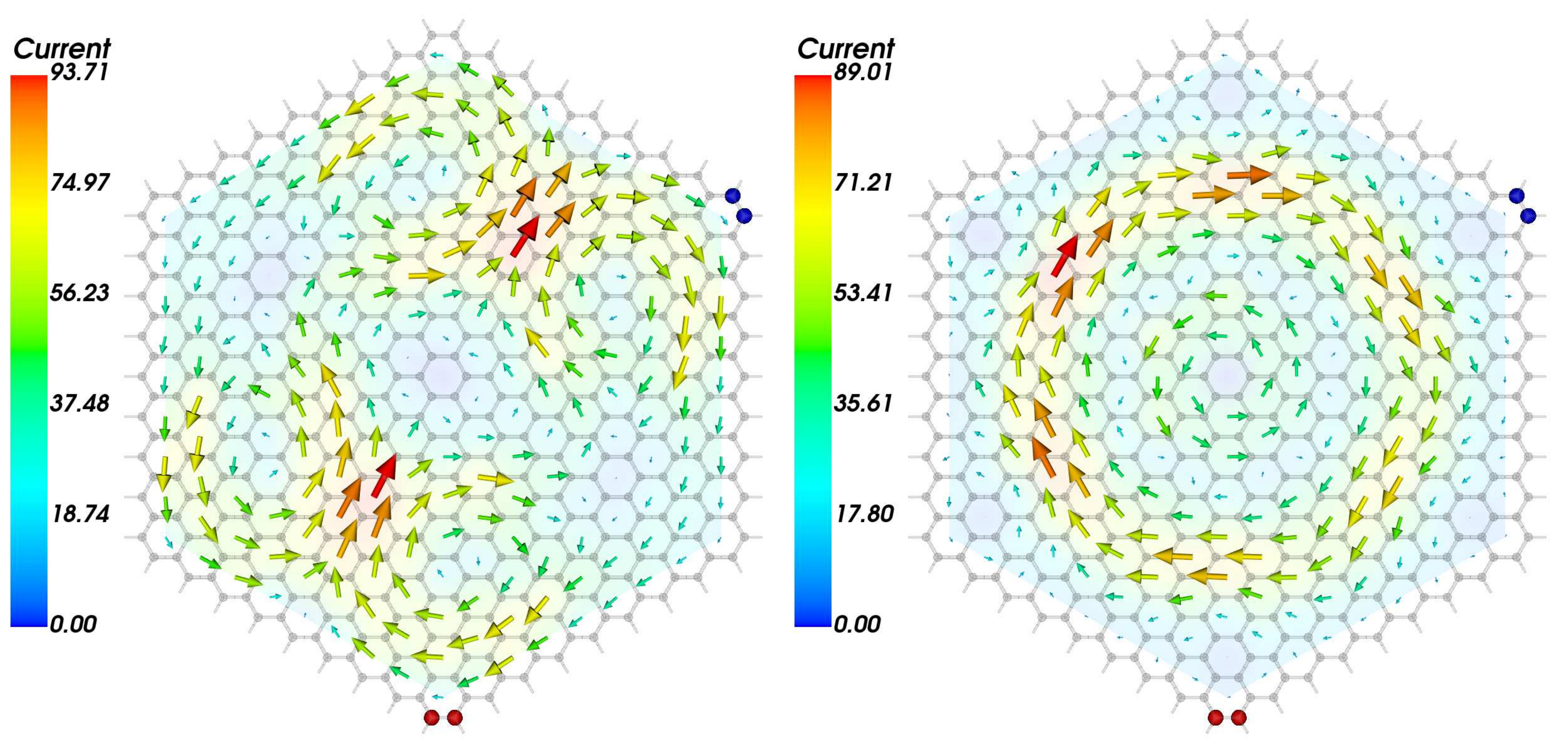}
	\caption{Multiple-vortex patterns for two-atom-wide-contact configurations. The energies of both left and right figures are $-1.3063eV$ and $-1.6680eV$, respectively. These energies are outside the energy range showed in the inset of Fig. \ref{dispersio+DOS}, and still  form interesting current patterns. The color bar presents a scale for the current intensity in units of $\frac{2e}{h}$. Contacts are represented by colored dots: source (red) and drain (blue).}
	\label{maxvsnu1}
\end{figure}
%essa é a 11 original
%\begin{figure}[htbp]
%    \centering
%    \includegraphics[width=0.45\textwidth]{current_0p6409n_0p3028n_0p1068_0p3808.pdf}
%    \caption{Top: transmission as a function of injection energy for contacts asymmetrically placed on the zigzag edge of the hexagonal GQD. Bottom: current patterns corresponding to the selected transmission peaks in the sequence a, b, c, d, from left to right, from top to bottom. The color bar presents a scale for the current intensity in units of $\frac{2e}{h}$. Contacts are represented by colored dots: source (red) and drain (blue). } \label{2zigzag}
%\end{figure}
In all cases studied, the current patterns corresponding to transmission peaks near zero energy are dominated, not surprisingly, by the edge states. Although the edge states appear at zero energy in the free GQD (see Figs. \ref{dispersio+DOS} and \ref{three_ldos} Z), with the contacts and  the chemical potential difference between source and drain, the energy values where they occur are shifted from zero. For contacts placed directly in front of each other across the hexagonal GQD (see Fig. \ref{2atomfront}) there is no reason to expect  current vortices, since our model does not contemplate the hydrodynamic viscous regime and the overall mirror symmetry is respected by the current flow.  Indeed, we did not observe vortices in these cases. Nevertheless, when the contacts were placed at two non-frontal armchair-like vertices, which are the cases of Figs. \ref{vortfig} and \ref{vortfig2} on the GQD, vortices, as well as circulating edge currents, were easily found for different injection energies. Clearly, the vortices are a consequence of the combination of the distribution of energy states around the center of the GQD, the geometry of the contacts, and the boost given to the charge carriers by the potential difference between source and drain contacts.
In Fig. \ref{vortfig}, electrons were injected and collected, respectively, at 2-atom-wide contacts placed at vertices of the GQD. There, the atoms are in an armchair configuration. When injecting electrons at contacts on the zigzag edges, like those in Fig. \ref{2zigzag}, both the vortices and edge currents become substantially modified, as can be observed comparing  Figs. \ref{vortfig}c and \ref{2zigzag}c, and Figs. \ref{vortfig}a and \ref{2zigzag}a. The current patterns become more asymmetric and the maximum current is considerably reduced, suggesting a greater resistance at the zigzag contacts.   We emphasize that it may be particularly  tricky to place the contacts specifically at the GQD vertex in a real experiment. To the best of our knowledge it has not been done yet, however the production of hexagonal GQD \cite{simpson2002synthesis,egberts2014frictional,lee2019synthesis} as well as transport measurements through GQD \cite{guttinger2012transport} have already been achieved. 

\subsection{Symmetry effects}
As remarked by Waltz and coworkers \cite{walz2014current}, in their study of current patterns in a graphene nanoribbon, ''the flow has a pronounced tendency to form ring structures (eddies) with a local current strength that exceeds the (average) through current by orders of magnitude''. The regular hexagonal shape of our graphene flake leads the vortices to be concentric with  the hexagon. This  comes in very handy to control the placement of the magnetic flux associated to a given vortex.  Nevertheless, at high injection energies, more complex current patterns   appear involving multiple vortices, as shown in Fig. \ref{maxvsnu1}. 
When the source and drain are at centrosymmetric positions relative to each other across the sample, like in Fig. \ref{2atomfront}, the current distributes itself equally on both sides of the line joining the contacts. When this symmetry is broken, by placing the contacts at non-centrosymmetric positions, like in Figs \ref{vortfig}, \ref{vortfig2} and \ref{maxvsnu1}, the formation of vortices is favored.  In fact, as shown in Ref. \cite{walz2014current} the  symmetry forbids the current vortices but when it is broken, they should be generic encounters.  
The importance of the states at energies near the injection energy of the electrons is very clear in the case of the edge currents, which are highly localized. On the other hand, the bulk currents are global, in the sense that they extend over the sample. In this case, all states will contribute, but those near the given energy contribute the most. This behavior was observed \cite{stegmann2020current} in aromatic carbon molecules studied via NEGF in connection with Density Functional Theory and explained by a simple tight-binding model.
\subsection{Edge defects and disorder}
\begin{figure}[htbp]
    \centering
    \includegraphics[width=0.45\textwidth]{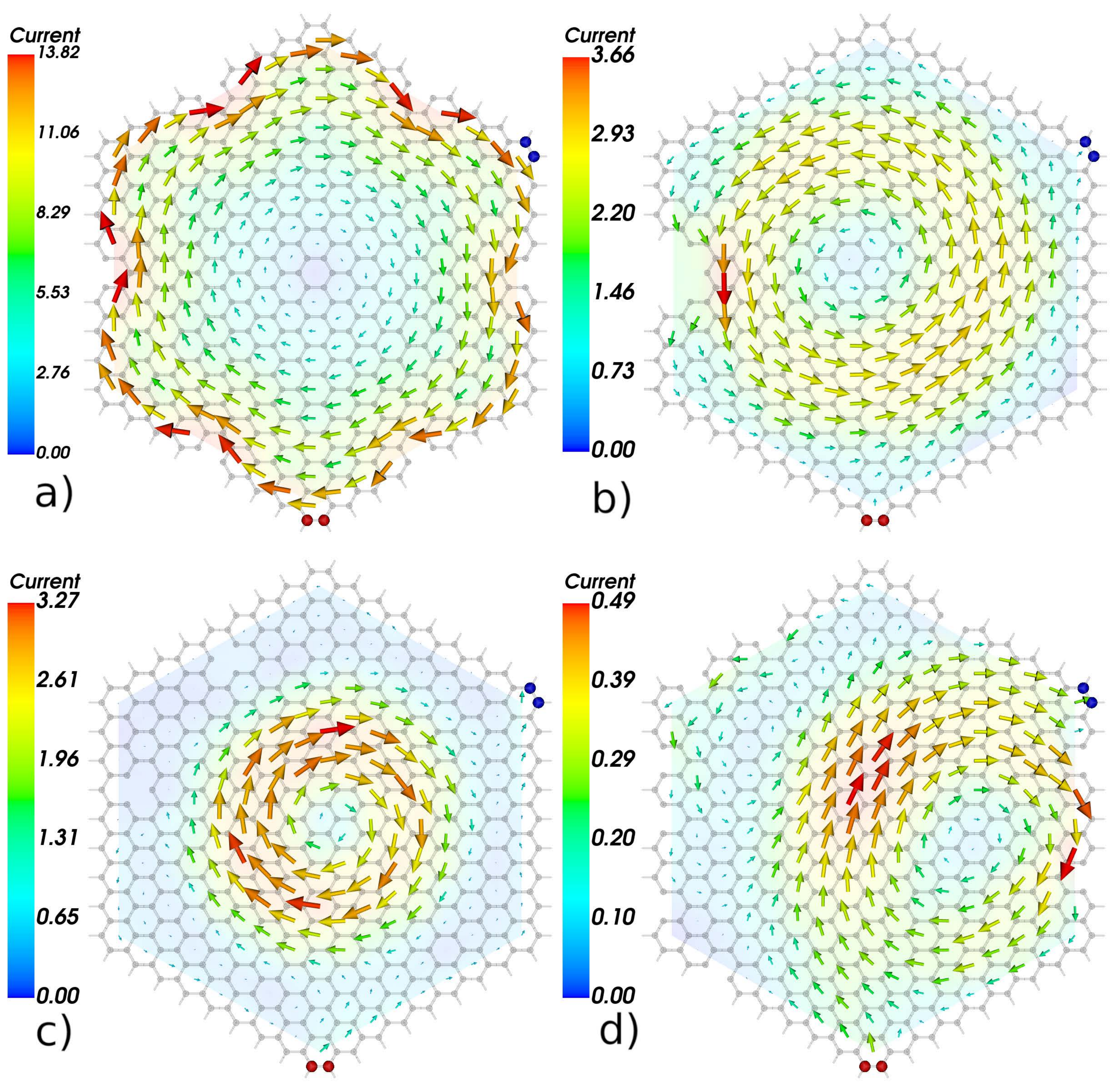}
    \caption{Current patterns for defected structures for energies $-0.3163eV, -0.6519eV, -0.9965eV, -0.6909eV$, respectively. The hydrogen bonds in the atoms near the defect were removed from the figure in order to improve visualization. The color bar presents a scale for the current intensity in units of $\frac{2e}{h}$. Contacts are represented by colored dots: source (red) and drain (blue). } \label{defects}
\end{figure}

As pointed out by Aharon-Steinberg and coworkers \cite{aharon2021long} in their study of nontopological edge currents in graphene,  if the current that flows on the edges is disrupted locally either by  disorder or by a tip potential, at zero magnetic field the current tends to bypass the disruption through the bulk and then return to the edge. In our simulations, when the energy is close enough to zero, such as in Figs. \ref{vortfig}d and \ref{vortfig2}b, the insertion of defects causes the current flowing in the edges to dodge the disruption. Incidentally, when symmetry is used to distribute the defects,  we can also see a circular flow as shown in Fig. \ref{defects}a. This can be explained by Bloch tunneling of the electrons along the periodic defect structure. However, if there is irregular-spaced disruption, disorder takes place and the constructive interference in which the circular flow is kept vanishes.
It is worth mentioning that the current shown in Fig. \ref{defects}a is not restricted to the edge, flowing through the bulk as well, something similar to what happens in Figs. \ref{vortfig}d and \ref{vortfig2}b. Edge currents, as shown in Figs. \ref{vortfig}c and \ref{vortfig2}c, were not observed in the presence of  defects, irrespective if they are placed on the edge or in the bulk. In fact, when there is a defect on the structure, the density of states for zero energy is not symmetrically distributed  over the edges as those shown in Fig. \ref{three_ldos}Z. The contact configuration of Fig. \ref{defects} is the same as  that of Fig. \ref{vortfig}, in order to have a clear comparison between the respective results.

Vortex-like patterns can still appear even if the defect is more pronounced, as in the case of Fig. \ref{defects}b. Indeed, the vortices can appear even when the defects are in the bulk (see Fig. \ref{defects}c). Interestingly, in Fig. \ref{defects}b the center of the vortex does not match the center of the structure, also, the local current is more intense near the defect, which may be responsible for the shifting of the vortex. 

When the defects are distributed randomly in the structure,  ``deformed vortices'' are formed, as in  Fig. \ref{defects}d. Notice that the maximum current of this deformed vortex is much smaller than  that of the perfectly circular pattern.   Also, when there is a defect, an exactly zero energy state appears both in the density of states and in the transmission, but, the respective current pattern does not show any vortex or circular flow, since it is a localized state. In fact, the effect of a single vacancy on the edges of zigzag graphene nanoribbons can change the  characteristic of the  edges from metallic to  semiconducting, due to localization.  See \cite{areshkin2007ballistic,poljak2012influence} for more information on the effect of disorder and vacancies in graphene nanoribbons.

\section{Concluding remarks}
The results presented here testify to the important role of contact geometry in the electronic properties of graphene quantum dots. When the position of the contacts are placed at centrosymmetric positions across the GQD, the current follows a direct path between the source and drain contacts. The path broadens and approaches the edge due to the increased importance of the edge states when the injection energy nears zero.  Current vortices   and circular edge currents appear when the centrosymmetry is broken and the contacts are placed in non-frontal armchair-like vertices on the GQD. Also, in the presence of the disorder introduced by defects, the edge currents do not appear. However, the bulk current that forms vortices seems to be robust even when there are  few defects. We point out that, when one has several defects, either on the edges or in the bulk, the edge circular current is destroyed and the intensity of the current of bulk vortices that may be formed is substantially reduced.  But, when the defects are distributed periodically over the edge, the edge current is maintained by Bloch tunneling.  Such  current vortices can be used in the design of quantum devices \cite{cir_cur_devices}, since their currents give rise to an extremely concentrated magnetic field. These currents  can also be used to generate spin-based quantum devices, since the orientation of a spin placed at the center of the vortex can be manipulated by the induced magnetic field \cite{circulating_fractal}.  Conversely, the current vortex will be affected by external magnetic perturbations in its proximity.   This suggests the use of  GQD, both as nanomagnets and as magnetic nanosensors. These might have applications in nanoscale magnetic recording or in electron microscopy, for instance.  Another possible application is  in the design of energy-tunable electron \cite{gerhard2012graphene} or even plasmonic \cite{smirnova2016trapping} lenses. Another interesting related subject is the study of magnetic textures associated to the multivortices. The study of  current vortices in differently shaped quantum dots is currently being performed in our group, including  both geometrical and topological deformations. We can anticipate that other geometries and shapes sustain such  currents as well, which increases the possibility of experimental verification of these results.

\section{Acknowledgments}
This work was supported by the Brazilian agencies Funda\c{c}\~ao de Amparo \`a Ci\^encia e Tecnologia do Estado de Pernambuco (FACEPE), Grant No. BIC-1187-1.05/20 (E.G.), and Conselho Nacional de Desenvolvimento Cient\'ifico e Tecnol\'ogico (CNPq), Grant No. 307687/2017-1 (F.M.). We also thank Dr. Denis R. Candido for invaluable discussions.

%\begin{figure}[htb!]
% \centering
% \includegraphics[scale=0.5]{dos1.png}
% \caption{\label{dos1}}  
%\end{figure}

%\begin{figure}[htb!]
% \centering
% \includegraphics[scale=0.5]{dos2.png}
% \caption{\label{dos2}}  
%\end{figure}

%\begin{figure}[htb!]
% \centering
% \includegraphics[scale=0.5]{dos3.png}
% \caption{\label{dos3}}  
%\end{figure}

\bibliographystyle{apsrev4-2}
\bibliography{ref}

\end{document}